\documentclass[reqno,12pt]{article}
\usepackage{amsmath,amsfonts,amssymb,amsthm,amstext,amscd,eucal}
\usepackage{graphics}
\usepackage{cite}
\usepackage{hyperref}
\usepackage{srcltx}
\usepackage[all]{xy}
\newcommand{\be}{\begin{equation}}
\newcommand{\ee}{\end{equation}}

\newcommand{\bse}{\begin{subequations}}
\newcommand{\ese}{\end{subequations}}
\newcommand{\bea}{\begin{eqnarray}}
\newcommand{\eea}{\end{eqnarray}}
\newcommand{\ba}{\begin{array}}
\newcommand{\ea}{\end{array}}

\makeatletter \@addtoreset{equation}{section}

\makeatother
\def\<{\langle}
\def\>{\rangle}

\begin{document}
\baselineskip 18pt%

\begin{titlepage}
\vspace*{1mm}%
\hfill%
\vbox{
    \halign{#\hfil \cr
%\;\;\;\;\;\;\;\;\;\;\;\;\; HIP-2009-nn/TH\cr
%           \;\;\;\;\;\;\;\;\;\;\;\;\;\;IPM/P-2009/024 \cr
%arXiv:090m.nnnn {\tt [hep-th]} \cr
           \cr
           } % end of \halign
      }  % end of \vbox
%\vspace*{7mm}%
%\vspace*{3mm}
\begin{center}
{{\Large {\bf  \textsl{CPT Violation Does Not Lead to Violation of Lorentz Invariance and Vice Versa }
}}}%
\vspace*{10mm}

%\vspace*{7mm}

{\bf \large{ Masud Chaichian$^a$, Alexander D. Dolgov$^{b,c,d}$, Victor A. Novikov$^d$\\ and
Anca Tureanu$^a$}}

\end{center}
\begin{center}
\vspace*{0.4cm} {\it { $^a$Department of Physics, University of Helsinki, P.O.Box 64, FIN-00014 Helsinki,
Finland\\
$^b$ Dipartimento di Fisica, Universit\`a degli Studi di Ferrara, I-44100
Ferrara, Italy\\
{$^c$} Instituto Nazionale di Fisica Nucleare, Sezione di Ferrara, I-44100
Ferrara, Italy\\
{$^d$}Institute of Theoretical and Experimental Physics, 113259 Moscow,
    Russia}}

\vspace*{.3cm}
({\it Phys. Lett. B}, in press)

\vspace*{.5cm}
\end{center}

\begin{center}{\bf Abstract}\end{center}
\begin{quote}
We present a class of interacting nonlocal quantum field theories, in
which the $CPT$ invariance is violated while the Lorentz invariance is
present. This result rules out a previous claim in the literature that
the $CPT$ violation implies the violation of Lorentz invariance.
Furthermore, there exists the reciprocal of this theorem, namely that
the violation of Lorentz invariance does not lead to the $CPT$ violation,
provided that the residual symmetry of Lorentz invariance admits the
proper representation theory for the particles. The latter occurs in the
case of quantum field theories on a noncommutative space-time, which in place of the broken Lorentz symmetry possesses the twisted Poincar\'e invariance. With such a $CPT$-violating interaction and the addition of a $C$-violating (e.g., electroweak) interaction, the quantum corrections due to the combined interactions could  lead to different properties for the particle and antiparticle, including their masses.
\end{quote}
%\vspace*{0.5cm}{\underline{\today}}

\end{titlepage}
\textwidth 16.5cm%

%\maketitle
%\setcounter{footnote}{0}
%\tableofcontents

\section{Introduction}

Lorentz symmetry and the $CPT$ invariance are two of the most fundamental
symmetries of Nature, whose violation has not yet been observed.
While the Lorentz invariance is a continuous symmetry of space-time, the $CPT$
involves the discrete space- and time-inversions, $P$, $T$, and the charge
conjugation operation on the fields, $C$. Although the  individual symmetries, $C$, $P$ and $T$
have been observed to be violated  in various interactions, their combined product,
$CPT$, remarkably remains still as an exact symmetry. The first proof of $CPT$
theorem was given by L\"uders and Pauli \cite{Luders-CPT,Pauli-CPT} based on the Hamiltonian
formulation of quantum field theory, which involves locality of the interaction,
Lorentz invariance and Hermiticity of the Hamiltonian. Later on the theorem was proven
by Jost \cite{Jost1} (see also \cite{Jost2,Streater,Bogoliubov}) within the axiomatic formulation of
quantum field theory without reference to any specific
form of interaction. This proof of $CPT$ theorem relaxes the requirement of locality
or "local commutativity"
condition to the so-called "weak local commutativity". Lorentz symmetry has been an essential ingredient
of the proof, both in the Hamiltonian and  in the axiomatic proofs.

A simple phenomenological classification of possible $C,\ P,\ T,\ CP,\ PT,\
TC$ and $CPT$-violating  effects is presented in \cite{Okun}.
For consequences of $CPT$ and their experimental tests, as well as some theoretical considerations on the
possibilities of violation
of Lorentz invariance and  $CPT$ in the known interactions, we refer
to \cite{Coleman,Colladay,Ellis,Kostelecky2,Cohen,Lehnert} and references therein.

 It is important to clarify the relation between the $CPT$ and Lorentz
invariance and in particular to see whether the violation of any of them
implies the violation of the other. This issue has recently become a topical
one  due to
the growing phenomenological importance of $CPT$ violating scenarios,
namely in neutrino physics as
well as  its cosmological and astrophysical consequences. Indeed, the relation
between the $CPT$ and Lorentz invariance has acquired a prominent place
in nowadays particle physics with the attempts of explaining in
a unified manner the contradictory
results, "anomalies", in the interpretation of various neutrino physics experiments, without
enlarging the neutrino sector. The idea was first suggested by Murayama and Yanagida \cite{Murayama}
in the form of different
masses for neutrino
and antineutrino, based on phenomenological considerations. This proposal was
formalized as a
$CPT$-violating quantum field theory with a mass difference between neutrino
and antineutrino in \cite{Barenboim1} (see also \cite{Bilenky}).
 The issue was taken up in relation with the Lorentz symmetry by Greenberg \cite{Greenberg},
the conclusion of Greenberg's analysis being
that $CPT$ violation implies violation of Lorentz invariance.
This result was given as a "theorem", the dispute on the validity  of which is the
subject of this Letter.

We should emphasize that a theorem which states that $CPT$
violation implies violation of Lorentz invariance has to
be explicit, first of all about what is meant by the charge conjugation in a
Lorentz violating
theory. Is the violation complete or is any subgroup of Lorentz symmetry left, which should have the
needed spin-representations to which the particles are assigned? Does the corresponding
theory which violates both $CPT$ and Lorentz invariance contain fields
with a plausible description in terms of equations of motion?

\section{$CPT$-violating free field model}

 A free field model in which particle and antiparticle have different
masses was proposed
in \cite{Barenboim1}.
Although the model was hoped to be Lorentz-invariant, a closer examination \cite{Greenberg} showed that
it is not -- the propagator is not Lorentz covariant, unless
the masses of particle and
antiparticle coincide. The model is also nonlocal and acausal:
the $\Delta(x,y)$-function, i.e. the commutator of two fields, does not
vanish for space-like separation, unless the two masses are the same, thus violating the Lorentz
invariance. This was
considered in \cite{Greenberg} as supporting
a general "theorem" that interacting fields that violate $CPT$
symmetry necessarily violate Lorentz invariance.

We would like to point out that the model taken in \cite{Greenberg} is utmost
pathological and can not be
considered as a quantum field theory. There, the claim was that the
model represents a free
complex scalar field, quantized in such a way that the mass
of the antiparticle differs from that of the particle. However, there is no definite equation of motion that this "field" satisfies, and no quantization procedure that would support the claim that the mode expansion with different masses for "particle" and "antiparticle" really represents a free quantized field. Also, two such "free fields" separated by a space-like
distance do not commute, i.e. the theory is acausal at the free level without invoking interaction.

Moreover, by requiring  that the classical symmetries and in particular
the global $U(1)$ symmetry for a free complex scalar field, i.e. the
conservation of electric charge, be preserved at the quantum level,
one can show that using the expansion for a free "field" as proposed in \cite{Greenberg}, would bring it back uniquely to the usual field expansion in
terms of creation and annihilation operators with $m= \bar{m}$ -- otherwise,
the electric charge is not conserved.

Furthermore, in a quantum field theory with acausal free fields, as taken in \cite{Greenberg}, observables,  which are functions of those fields,
do not commute when separated by space-like distances.  This, according to Pauli's proof of the spin-statistics theorem, implies that there is no spin-statistics relation already for {\it the free fields}. Thus, one has no rule whether to apply commutation or anticommutation relations in quantizing the fields. But the worst is that in such a model, where Lorentz invariance is violated by the free fields, there is {\it no concept of spin} to start with altogether.

\section{$CPT$-violating but Lorentz-invariant nonlocal model}

Here, as an example, we propose a model
which preserves Lorentz invariance while breaking the $CPT$ symmetry through a
(nonlocal) interaction. The latter attitude is taken as responsible for the violations
of a symmetry, based
on our experience that all the discrete, $C$, $P$ and $T$ invariances, as well as other symmetries,
are broken in our description of Nature by means of interaction. We also know that nonlocal field theories appear, in general, as effective field theories of a larger theory.

Consider a field theory with the nonlocal interaction Hamiltonian of the type
\be
{\cal H}_{int}(x)= \lambda\int d^4 y\ \phi^*(x)\phi(x)\phi^*(x) \theta(x_0-y_0)
\theta((x-y)^2) \phi (y) + h.c.,                       \label{(3.1)}
\ee
where $\lambda$ is a coupling constant with dimension appropriate for the Hamiltonian density, $\phi(x)$ is a Lorentz-scalar field in the interaction picture and $\theta$ is the Heaviside
step function, with
values 0 or  1, for its negative and positive argument, respectively.
The combination $\theta (x_0-y_0) \theta ((x-y)^2)$ in \eqref{(3.1)}  ensures the Lorentz invariance,
i.e. invariance under the
proper orthochronous Lorentz transformations, since the order of the times $x_0$ and $y_0$ remains
unchanged for time-like intervals, while for space-like distances the interaction vanishes.
Also, the
same combination makes the nonlocal interaction causal at the tree level, which dictates
that there is no interaction when the fields
are separated by space-like distances and thus there is a maximum
speed of $c=1$  for the propagation of information.

On the other hand, it is clear that $C$ and $P$ invariance are trivially satisfied in \eqref{(3.1)}, while
$T$ invariance
is broken due to the presence of $\theta(x_0-y_0)$ in the integrand.

One can always
insert
into the Hamiltonian \eqref{(3.1)}, without changing its symmetry properties, a weight function or
form-factor $F((x-y)^2)$, for instance of a Gaussian type:

\be
F= \exp\left(-\frac{(x-y)^2}{l^2}\right),                    \label{(3.2)}
\ee
with $l$ being a nonlocality length in the considered
theory. Such a weight function would smear out the interaction and would guarantee the
desired behaviour of the integrand in \eqref{(3.1)}; in the limit of
fundamental length $l \to 0$ in \eqref{(3.2)}, the Hamiltonian \eqref{(3.1)} would correspond to a local, $CPT$- and
Lorentz-invariant theory. A weight function such as \eqref{(3.2)} would make the acausality of the model (see the next section) restricted only to very small distances, of the order of $l$. The latter could be looked upon as being a characteristic parameter relating the effective field theory to its parent one, for instance the radius of a compactified dimension when the parent theory is a higher-dimensional one. Furthermore, with such a weight function, the interaction vanishes at infinite $(x-y)^2$ separations and thus one can envisage the existence of in- and out-fields.

There exists a whole class of such $CPT$-violating, Lorentz-invariant field
theories involving different, scalar, spinor or higher-spin interacting
fields. Typical simplest examples are:
\bea
\mathcal H_{int}(x)&=& \lambda\int d^4y\, \phi_1^*(x)\phi_1(x) \theta(x_0-y_0)
\theta((x-y)^2) \phi_2(y) + h.c.,   \label{(3.3)}\\
\mathcal H_{int}(x)&=& \lambda\int d^4y \,\bar\psi(x)\psi(x)\theta(x_0-y_0)
\theta((x-y)^2) \phi(y) + h.c.,\label{(3.4)}\\
\mathcal H_{int}(x)&=&\lambda\int d^4y\, \phi(x) \theta(x_0-y_0) \theta((x-y)^2) \phi^2(y) + h.c.                               \label{(3.5)}
\eea

\section{Quantum theory of such nonlocal interactions}

The $S$-matrix in the interaction picture is obtained as solution of the Lorentz-covariant Tomonaga-
Schwinger equation \cite{Tomonaga,Schwinger} (see also \cite{Schweber,Nishijima}):
\be i\frac{\delta}{\delta \sigma(x)} \Psi[\sigma] =
\mathcal H_{int}(x)\Psi[\sigma]\,,   \label{(4.1)}\ee
with $\sigma$ a space-like hypersurface, and the boundary condition:
\be\Psi[\sigma_0]=\Psi\,,\label{(4.2)}\ee
where $\mathcal H_{int}$ is for instance the Hamiltonian \eqref{(3.5)}
with the fields in the interaction picture.
Then Eq. \eqref{(4.1)} with the boundary condition \eqref{(4.2)} represent a well-posed
Cauchy problem.

The existence of a unique solution for the Tomonaga-Schwinger equation
is ensured if the integrability condition
\be
\frac{\delta^2 \Psi[\sigma]}{\delta \sigma(x) \delta \sigma(x')}
- \frac{\delta^2 \Psi[\sigma]}{\delta \sigma(x') \delta \sigma(x)}=0,       \label{(4.3)}
\ee
with $x$ and $x'$ on the surface  $\sigma$, is satisfied. The integrability
condition \eqref{(4.3)}, inserted into \eqref{(4.1)}, requires that the commutator
of the interaction Hamiltonian densities vanishes at space-like separation:
\be\label{(4.4)}
[{\cal H}_{int}(x),{\cal H}_{int}(y)]=0\,,\  \quad \text{for}\  (x-y)^2<0\,.
\ee

Since in the interaction picture the field operators satisfy free-field
equations, they automatically
satisfy Lorentz-invariant commutation rules. The Lorentz-invariant
commutation
relations are such that \eqref{(4.4)} is fulfilled only when $x$ and $y$ are
space-like separated,
$(x - y)^2 < 0$ ,
i.e. when $\sigma$ is a space-like surface. As a result, the integrability
condition \eqref{(4.4)} is equivalent to
the microcausality condition for local relativistic QFT. When the
surfaces $\sigma$ are hyperplanes
of constant time, the Tomonaga-Schwinger equation reduce to the
single-time Schr\"odinger
equation.

Inserting, e.g., the expression \eqref{(3.5)} into \eqref{(4.4)}, we have:
\bea\label{(4.5)}
[{\cal H}_{int}(x),{\cal H}_{int}(y)]=&\lambda^2\int d^4a\, d^4 b\  \theta((x-a)^2)\theta(x^0-a^0)\theta((y-b)^2)\theta(y^0-b^0)\cr
&\times[\phi(x)\phi^2(a) + h.c.,\phi(y)\phi^2(b) + h.c.]\,.
\eea
The commutator on the r.h.s. will open up into a sum of products of field at the points $x,y,a,b$, multiplied by commutators of free fields like $[\phi(x),\phi(y)]$, $[\phi(x),\phi(b)]$, $[\phi(a),\phi(y)]$, $[\phi(a),\phi(b)]$. In order for the commutator \eqref{(4.5)} to vanish, all the coefficients of the products of fields in the expansion have to vanish, since the fields at different space-time points are independent. Clearly, the terms with the coefficient $\Delta(x-y)=[\phi(x),\phi(y)]$ vanish for $(x-y)^2<0$. However, the commutator \eqref{(4.5)} does not vanish for $(x-y)^2<0$. In order to show this, it is enough to show that one independent product of fields has nonzero coefficient.

Let us consider the products which contain the fields $\phi(x), \phi(y), \phi(a),\phi(b)$. A straightforward calculation shows that the terms containing these fields are:
\be
\int d^4a\, d^4 b\  \theta((x-a)^2)\theta(x^0-a^0)\theta((y-b)^2)\theta(y^0-b^0) 2\Delta(a-b)\{\phi(a),\phi(b)\}\phi(x)\phi(y)+h.c.\label{(4.6)}
\ee
A closer study of the expression \eqref{(4.6)} shows that it does not vanish at space-like distances between $x$ and $y$ and thus the causality condition \eqref{(4.4)} is not
satisfied.

This, in turn, implies that the  field operators in the Heisenberg picture, $\Phi_H(x)$ and $\Phi_H(y)$, do not satisfy the locality condition
  \be
          [\Phi_H(x), \Phi_H(y)]=0,\ \quad   \text{for}\  (x-y)^2<0,   \label{(4.7)}
\ee
when the quantum corrections are taken into account. This is in accord with the requirement of locality condition \eqref{(4.7)} for the validity of $CPT$ theorem both in the Hamiltonian proof \cite{Luders-CPT,Pauli-CPT} and as well in the axiomatic
one \cite{Jost1}-\cite{Bogoliubov}, taking into account that there is no example of a QFT, which satisfies the weak local commutativity condition (WLC) but not the local commutativity (LC).
For general considerations on the causality and unitarity properties of nonlocal relativistic quantum field theories,
we refer to \cite{Marnelius1,Marnelius2} and references therein.

Instead of the description in terms of $\mathcal H_{int}$ and the interaction picture as done above, one can also consider a whole class of CPT-violating, Lorentz-invariant nonlocal quantum field theories described by their actions or Lagrangians. An example, analogous to \eqref{(3.5)} is given by the following action:
\begin{equation}
S=\int d^4x \left(\frac{1}{2}{\partial_\mu \Phi_H(x) \partial^\mu \Phi_H(x)- \frac{1}{2}m^2 \Phi_H^2(x)- \lambda \int d^4y \left(\Phi_H(x) \theta(x_0-y_0) \theta((x-y)^2)\Phi_H^2(y)+h.c.\right)}\right),      \label{4.8}
\end{equation}
with the corresponding field equation given by
\begin{equation}
( \Box + m^2) \Phi_H(x)=-\lambda \int d^4y\, \theta(x_0-y_0)
 \theta((x-y)^2)\left( \Phi_H^2(y)+ 2 \Phi_H(x) \Phi_H(y)+ h.c.\right).           \label{4.9}
\end{equation}
  Analogous to \eqref{(3.1)}-\eqref{(3.4)}, nonlocal actions can be written down in a similar way.

   We recall that the relation between the action (or Lagrangian) and the Hamiltonian in a nonlocal relativistic field theory is not so straightforward as in the case of local field theories. For instance, from the action \eqref{4.8} does not follow the Hamiltonian given  by \eqref{(3.5)} and one should adopt instead a more involved prescription (see, e.g., \cite{Marnelius1,Marnelius2}). The quantum treatment of such theories as well should  be performed through the use of Yang-Feldman equation \cite{Yang} with the fields, denoted by $\Phi_H(x)$, in the  Heisenberg picture.

With such a $CPT$-violating interaction as in \eqref{(3.1)}-\eqref{(3.5)} or \eqref{4.8}, and the addition of a $C$-violating (e.g., electroweak) interaction, the quantum corrections due to the combined interactions could  lead to different properties for the particle and antiparticle, including their masses.

\section{Lorentz-invariance violating but $CPT$-invariant quantum field theories:
Reciprocal theorem}

During the last decade, we have learned  that the violation of Lorentz invariance does not necessarily lead to the
violation of the $CPT$ theorem. The example comes from the  quantum field theory on noncommutative space-time (NC QFT) with the canonical, Heisenberg-like, commutation relations for coordinate operators:
\be
                   [x^\mu, x^\nu]=i \theta^{\mu\nu},       \label{(5.1)}
\ee
with $\theta^{\mu\nu}$ an antisymmetric constant matrix \cite{SW}.

In this case,
by the nature of the above noncommutativity parameter $\theta^{\mu\nu}$ being a constant but not a tensor, Lorentz invariance is broken, but not the  $CPT$
symmetry \cite{CNT,LAG,CMNTV,SSJ-CPT}. Translational invariance is valid. In addition to the Lorentz invariance violation, such NC QFTs are nonlocal in
the noncommuting coordinates. However, the Lorentz symmetry violation is of a very particular
form, and invariance under the stability group of the matrix $\theta^{\mu\nu}$ is preserved under the so-called residual
symmetry $O(1,1)\times SO(2)$ \cite{LAG-B-Z}. This reduced symmetry is enough to prove the $CPT$ theorem only for the scalar
fields (for which the $C$ operation is a simple Hermitian conjugation) on the noncommutative space-time \eqref{(5.1)}
\cite{LAG}.
A full proof of the $CPT$ theorem in Lorentz-violating noncommutative quantum field theory, however, could be achieved \cite{CMNTV} only by using the twisted Poincar\'e symmetry \cite{CKNT,CPrT} which these theories possess. The twisted Poincar\'e invariance is a deformation of the Poincar\'e symmetry, considered as a Hopf algebra, a concept coming from the theory of quantum groups \cite{Chai-Dem}, as compared with the Lie algebra. The
irreducible representations of twisted Poincar\'e are identical to those of the usual
Poincar\'e algebra, i.e. labeled by the mass and spin of the particles \cite{CKNT,CPrT}.
Therefore, the meaning of the charge conjugation has survived intact in the noncommutative quantum field
theories.
While parity and time reversal symmetries can be defined with any concept of space and
time, the notion of charge conjugation has meaning only in the framework of Lorentz symmetry.
Antiparticles are a consequence of special relativity. Particle and antiparticle are in the same
irreducible representation of the Poincar\'e group. The $CPT$ theorem is thus strongly connected to
the Poincar\'e group representations, and not so much to the Lorentz symmetry, as the validity of the
$CPT$ theorem in the noncommutative space-time shows.

There are other examples of Lorentz-invariance violating but
CPT-invariant theories, as in the extensions of the Standard Model given
in \cite{Colladay} or with aether compactification \cite{Carroll}. However, in such
cases the Lorentz invariance broken theory does not in general admit the usual representation content for the particles, unless the breaking of Lorentz invariance is made to be a spontaneous one.

\section{Conclusion}

We have presented a whole class of interacting nonlocal quantum field theories, such as the ones
in \eqref{(3.1)}-\eqref{(3.5)} or \eqref{4.8}, which violate $CPT$ invariance while being Lorentz-invariant. This result
invalidates a general claim made previously \cite{Greenberg}, that "$CPT$ violation implies violation of Lorentz invariance".
With such a CPT-violating interaction as in \eqref{(3.1)}-\eqref{(3.5)} or \eqref{4.8}, and the addition of a C-violating (e.g., electroweak) interaction, the quantum corrections due to the combined interactions could  lead to different properties for the particle and antiparticle, including their masses. Furthermore, there exists the reciprocal of this theorem, namely that the violation of Lorentz invariance does not necessarily lead to $CPT$ violation.

\section*{Acknowledgements}

We are grateful to Luis \'Alvarez-Gaum\'e, whose belief in the result of this Letter
encouraged
us to pursue the study. We are indebted to Jos\'e Gracia-Bond\'ia
and Peter Pre\v{s}najder for useful discussions and for their invaluable contribution
to this work. We thank Samoil Bilenky for many clarifying discussions on neutrino-antineutrino data in connection with the $CPT$ violation. A.D.D. and V.A.N. would like to thank the RF Ministry for Science and Education for partial support under the Contract 02.740.11.5158. A.T. acknowledges the support of the Academy
of Finland under the Project Nos. 136539 and 140886, and the support for this research of the Vilho, Yrj\"o and Kalle V\"ais\"al\"a Foundation, Finland.

%%%%%%%%%%%%%%%%%%%%%%%%%%%%%

\end{document}